# Entanglement Between Stochastic Atomic Motion and Three-Level Atom

Ahmed Salah[1], M. A. El-Sayed[1] and N. H. Abdel Wahab[2],

[1]Mathematics and Theoretical Physics Department, Nuclear Research Center, Egyptian Atomic Energy Authority, P. No. 13759, Cairo, Egypt.
[2]Mathematics Department, Faculty of Science, Minia University, Minia, Egypt

**Abstract**

In this paper, we are interested in studying entropy and dynamics entanglement between a single time dependent three-level atom interacting with one-mode cavity field when the atomic motion is taken into account. An exact analytical solution for the wave function is given by using Schrödinger equation for a specific initial condition. The field entropy of this system is investigated in the non-resonant case. The influences of the detuning parameters and atomic motion on the entanglement degree are examined. We show that both of the detuning parameters and atomic motion play important roles in the evolution of von Neumann entropy and atomic populations. Finally, conclusion and some features are given.

## 1. Introduction

The Jaynes-Cummings model (JCM) [1] in the rotating wave approximation has been a subject of intense investigations [2–5]. Recently, increased attention has been paid to the atomic coherent states [6]. It was found that the JCM initially in an atomic coherent state together with the vacuum field can generate field squeezing as well as squeezing of the fluctuations of the atomic dipole variables [3–5] and it has been shown that squeezed atoms can radiate squeezed light [3,4]. The relationship between the field and atomic squeezing in the thermal JCM with an initially coherent atom has been discussed by a few authors [4, 7]. The problem of interaction of matter with squeezed light has been extensively studied for the past ten years [8–10]. However, the influence of nonlinearity in one-photon process on the squeezing properties of the radiation field and the fluctuations of the atomic dipole variables has not been studied earlier. Nonlinear one-photon processes are important for understanding the generation of squeezed states in off-resonant fluorescence [11].

This model was generalized in many directions, such as adding further levels [12] where the three-level ($\Lambda$, V and $\Xi$-type) atomic systems have been discussed [13–16]. Moreover, the effects of the atomic motion [17–20], the Kerr medium [21] and the phases of the field and of the atom [22, 23] are demonstrated

Recently, much more attention has been paid to the von Neumann entropy in the presence of the atomic motion. For instance, the effects of atomic motion and field-mode structure in the Jaynes– Cummings model have been studied in [24, 25]. It has been shown that the atomic motion leads to a periodic evolution of the von Neumann entropy. Also, these results have been extended to studying the mixed-state entanglement and analysis of the pattern entanglement between a three-level trapped ion and the laser field [26]. Another extension consists in examining the influences of the atomic motion and field-mode structure on the dynamical properties of the Wehrl entropy of the cavity field and comparing it with the von Neumann entropy for two-photon processes [27]. Furthermore, the time evolution of the field (atomic) entropy reflects the time evolution of the degree of entanglement between the atom and the field. An expression for the field entropy for the entangled state of a single two-level atom interacting with a single electromagnetic field mode in an ideal cavity with the atom undergoing either a one or a two-photon transition has been studied [28]. Also, much attention has been focused on the properties of the entropy and phase of the field in the Jaynes-Cummings model due to the entropy theory of the interaction of the field with an atom presented by Phoenix and Knight (PK) [29-30].

In this paper, we study the interaction between an atomic system described by a three-level $\Lambda$-type atom and a quantized radiation field in the rotating wave approximation. The field has single model and interaction is affected by a two-photon process. The wave function of the atomic system is obtained. We show different phenomena related to this problem when the detuning parameter and the atomic motion are taken in to account. With numerical stimulation, the influence of both detuning parameters and atomic motion on the degree of entanglement are discussed when the atom is initially prepared in exited state and the field in a coherent state.

Our paper is organized as follows: In Sec.2 we present the model which describe a three-level lambda-type atom one-mode system. Also, the wave function of the considered atomic system is calculated. In Sec.3 we give the general formula for the entropy field. The numerical results are given in Sec.4. Finally, a summary and conclusion are illustrated.

## 2. Model and the Wave Function

We assume that we have a single three-level lambda-type atom with $\omega_j (j=1,2,3)$ and $\omega_1 > \omega_2 > \omega_3$ as the transition energy between three level atom levels. This atom is interacting with a single mode cavity field. The field mode is described by creation (annihilation) operator $\hat{a}^\dagger (\hat{a})$ and frequency $\Omega$. In the rotating wave approximation the total Hamiltonian $\hat{H}$ for considered system can be written in the form

$$\hat{H} = \hat{H}_0 + \hat{H}_{IN}, \qquad (1)$$

where $\hat{H}_0$ is the free part and $\hat{H}_I$ is the interacting part of the Hamiltonian. Let us divide $\hat{H}$ as follows

$$\hat{H}_0 = \sum_{j=1}^{3} \omega_j \sigma_{jj} + \Omega \hat{a}^\dagger \hat{a}, \qquad (2)$$

where $\sigma_{jj} = |i\rangle\langle j|$ are the population operators. On the other hand, we consider the interaction between the atomic system and the field to be affected through two-photon processes where the two photon needed to accomplish the. In the non-resonant case the interaction Hamiltonian of this system is given by

$$\hat{H}_{IN} = \sum_{s=1}^{2} \lambda_s g(z) \left[ \hat{a} e^{-i\Delta_s t} \sigma_{1,s+1} + h.c \right] \qquad (3)$$

where

$$\Delta_s = \Omega + \omega_{s+1} - \omega_1 \qquad (4)$$

where $\Delta_s$ is the detuning parameter. $\lambda_s$ is the atom-field coupling constant and $g(z)$ is the shape function of the cavity field mode. We restrict our study for the stochastic motion along z axis, in that case

$$g(z) \to g(vt) = \sin\left(\frac{p\pi vt}{L}\right) \qquad (5)$$

where $v$ denote the stochastic motion velocity, $p$ represent the number of half-wave length of the field mode inside the cavity of the length $L$. For a particular choice of the stochastic motion velocity $v = \lambda_s L / \pi$, thus $g(z) = \sin(p\lambda_s t)$. Now, we focus to solve the model of considered atomic system, thus we write the atom-field wave function of this system at an arbitrary time t in the form

$$|\psi(t)\rangle = \sum_{n=0}^{\infty} q_n \left[ \sum_{j=1}^{3} \psi_j(n,t) |n_j\rangle \otimes |j\rangle \right] \quad (6)$$

where $q_n = \exp(-\frac{\bar{n}}{2})\alpha^n / \sqrt{n!}$, with $\bar{n} = |\alpha|^2$ is the initial mean photon number for the model. However $\psi_j(n,t)$ are the probability amplitudes coefficients to be determined from the initial state $|\psi(t)\rangle$ and $|n_j\rangle$ are the Fock states of the mode field.

Using the timed-dependent Schrödinger equation $(i\partial/\partial t)|\psi(t)\rangle = \hat{H}_{IN}|\psi(t)\rangle$ and action of $\hat{a}$ and $\hat{a}^\dagger$ on the state $|n\rangle$, we obtain the following system of ordinary differential equations:

$$i\frac{d}{dt}\begin{pmatrix}\psi_1\\\psi_2\\\psi_3\end{pmatrix} = \begin{pmatrix}0 & F_1 & F_2\\F_1^* & 0 & 0\\F_2^* & 0 & 0\end{pmatrix}\begin{pmatrix}\psi_1\\\psi_2\\\psi_3\end{pmatrix} \quad (7)$$

where

$$F_s = f_s e^{-i\Delta_s t} \quad f_s = \lambda_s g(vt)\sqrt{n+1} \quad (8)$$

and $F_s$ is the conjugate of $F_j^*$. We turn our attention to solve the system, where the exact solution can be obtained where $\lambda_s g(vt) = \lambda_s$. Let us assume that at time $t=0$ the atom is initially prepared in the upper state $|1\rangle$. Thus, the initial wave function can be decomposed into its atomic and field part in the form:

$$|\psi(t=0)\rangle = \sum_{n=0}^{\infty} q_n |1,n\rangle \quad (9)$$

Let us firs start with the solution for this system in the resonant case $\Delta_1 = \Delta_2 = 0$. In this case, the probability amplitudes $\psi_j$ are given by

$$\psi_1(t) = \cos(\delta t), \quad \psi_2(t) = -if_1\frac{\sin(\delta t)}{\delta}, \quad \psi_3(t) = -if_2\frac{\sin(\delta t)}{\delta}. \quad (10)$$

where

$$\delta = \sqrt{f_1^2 + f_2^2}$$

Also, the probability amplitudes $\psi_j$ in the off-resonance case $\Delta_1 = \Delta_2 = \Delta$ are given by

$$\psi_1(t) = e^{i\Delta t/2}\left\{\cos(\eta t) - \frac{i\Delta}{\eta}\sin(\eta t)\right\},$$

$$\psi_2(t) = \frac{2f_1 e^{i\Delta t/2}}{\eta(\Delta^2 - \eta^2)}\left\{i\Delta\eta(e^{i\Delta t/2} - \cos(\eta t)) + (\Delta^2 - \eta^2)\sin(\eta t)\right\}, \quad (11)$$

$$\psi_3(t) = \frac{2f_2 e^{i\Delta t/2}}{\eta(\Delta^2 - \eta^2)}\left\{i\Delta\eta(e^{i\Delta t/2} - \cos(\eta t)) + (\Delta^2 - \eta^2)\sin(\eta t)\right\}$$

where

$$\eta = \sqrt{\left(\frac{\Delta}{2}\right)^2 + \delta^2}$$

Furthermore, the solution for this system in the in the non-resonant case $\Delta_1 \neq \Delta_2$ and after some algebra the probability amplitudes $\psi_j$ are

$$\begin{pmatrix} \psi_1 \\ \psi_2 \\ \psi_3 \end{pmatrix} = \begin{pmatrix} -\dfrac{1}{f_2} \sum_{j=1}^{3} C_j \mu_j e^{i(\mu_j - \Delta_2)t} \\ -\dfrac{1}{f_1 f_2} \sum_{j=1}^{3} C_j \left( \mu_j(\mu_j + \Delta_2) - f_2^2 \right) e^{i(\mu_j - \Delta_2 - \Delta_1)t} \\ \sum_{j=1}^{3} C_j e^{i\mu_j t} \end{pmatrix} \qquad (12)$$

where

$$C_j = \frac{\Delta_2 + \mu_k + \mu_l}{\mu_{jk} \mu_{jl}}, \qquad (13)$$

where $\mu_{jk} = \mu_j - \mu_k$ $i \neq j \neq k = 1,2,3$ and $\mu$ satisfy the third-order equation

$$\mu^3 + x_1 \mu^2 + x_2 \mu + x_3 = 0, \qquad (14)$$

where

$$\begin{aligned} x_1 &= \Delta_1 - 2\Delta_2, \\ x_2 &= \Delta_2(\Delta_1 - \Delta_2) - f_1^2 - f_2^2, \\ x_2 &= f_2^2(\Delta_2 - \Delta_1), \end{aligned} \qquad (15)$$

These roots are given by :

$$\mu_j = -\frac{1}{3} x_1 + \frac{2}{3} \sqrt{x_1^2 - 3x_2} \cos\left( \xi + \frac{2}{3}(j-1)\pi \right), \qquad (16)$$

with

$$\xi = \frac{1}{3} \cos^{-1}\left( \frac{9 x_1 x_2 - 2 x_1^3 - 27 x_3}{2\left(x_1^2 - 3x_2\right)^{3/2}} \right).$$

In the important to mention that the wave function of assumed atomic system, with the stochastic motion taken into account is obtained by computer simulation. Now, with the wave function given, we can discuss dynamics properties of considered system. In the following section, we shall investigate numerically the influence of the detuning parameters and the stochastic motion on the dynamics behavior of the von Neumann entropy and on the population. This will be done in the following section when we suppose the atom to be initially prepared in its upper state and the input field in a coherent state as in (9).

## 3. von Neumann Entropy

Quantum entropy was first formulated by von Neumann [29-30] as an extension of the Gibbs entropy in classical statistical mechanics. The entropy of a quantum system is given by the von Neumann entropy

$$S = -Tr\{\rho \ln \rho\}, \qquad (17)$$

It gives zero for all pure states, i.e., for states which satisfy the condition $\hat{\rho}^2 = \hat{\rho}$, where $\hat{\rho}$ is the density operator describing a given quantum state. For this reason, this entropy cannot distinguish between various pure states and it is rather the measure of purity of quantum states. Then von Neumann entropy is defined as

$$S_{A(F)} = Tr_{A(F)}\{\rho_{A(F)} \ln \rho_{A(F)}\} \qquad (18)$$

where the reduced density operator $\rho_{A(F)}$ are

$$\hat{\rho}_{A(F)} = Tr_{F(A)}\{\hat{\rho}_{A(F)}\} \qquad (19)$$

Since the trace is invariant under a similarity transformation, we can go to a basic in which the atomic density matrix is diagonal and write Eq. () as

$$S_F = S_A = -\sum_{i=1}^{3} \xi_i(t) \ln \xi_i(t) \tag{20}$$

where $\xi_i$ is the eigenvalue for the atomic density matrix is $\rho_A(t)$ expressed as Eq. (18). Furthermore, the eigenvalues $\xi_i$ are the roots of the cubic equation with one unknown

$$\xi^3 + A\xi^2 + B\xi + C = 0 \tag{21}$$

Here the coefficients A, B, C are given by the matrix elements of Eq. ():

$$A = -(\rho_{11} + \rho_{22} + \rho_{33}) = -1$$
$$B = \rho_{11}\rho_{22} + \rho_{11}\rho_{33} + \rho_{22}\rho_{33} - \rho_{13}\rho_{31} - \rho_{23}\rho_{32} - \rho_{12}\rho_{21} \tag{22}$$
$$C = \rho_{13}\rho_{31}\rho_{22} + \rho_{32}\rho_{23}\rho_{11} + \rho_{12}\rho_{21}\rho_{33} - \rho_{11}\rho_{22}\rho_{33} - \rho_{12}\rho_{23}\rho_{31} - \rho_{32}\rho_{21}\rho_{12}$$

Therefore, the eigenvalue $\xi_i$ is given as

$$\xi_1 = \frac{1}{3} - 2R\cos\left(\frac{\eta}{3}\right)$$
$$\xi_2 = \frac{1}{3} + 2R\cos\left(\frac{\eta}{3} + \frac{\pi}{3}\right) \tag{23}$$
$$\xi_3 = \frac{1}{3} + 2R\cos\left(\frac{\eta}{3} - \frac{\pi}{3}\right)$$

where

$$\eta = \cos^{-1}\frac{q}{R^3}$$
$$q = -\frac{1}{27} + \frac{1}{6}B + \frac{1}{2}C \tag{24}$$
$$R = \pm\left(\frac{1}{3}B - \frac{1}{9}\right)^{1/2}$$

In what follows, we shall investigate the influence of the detuning and the atomic on the behavior of both the von Neumann entropy and the population of the atomic system

## 4. Numerical Results and Discussion

In this section we shall study the influence of the detuning parameters and atomic motion on the population and on the entanglement measured by von Neumann entropy for the pervious considered system. The numerical results are obtained in Fig. 1-4. In our investigations, we assume the input field is in a coherent state and we chose the mean value number $\bar{n} = 25$. For simplicity, we consider the coupling constants have been taken real and equal $\lambda_1 = \lambda_2 = \lambda$. We plot both the population and von Neumann entropy against the scaled time. Furthermore, the scaled time is $\lambda t$ when the atomic motion is not taken in to account while is $(1-\cos(pt))/p$ when the atom is in motion. Also, in all figures, left curves cover the von Neumann entropy while the right curves cover the populations.

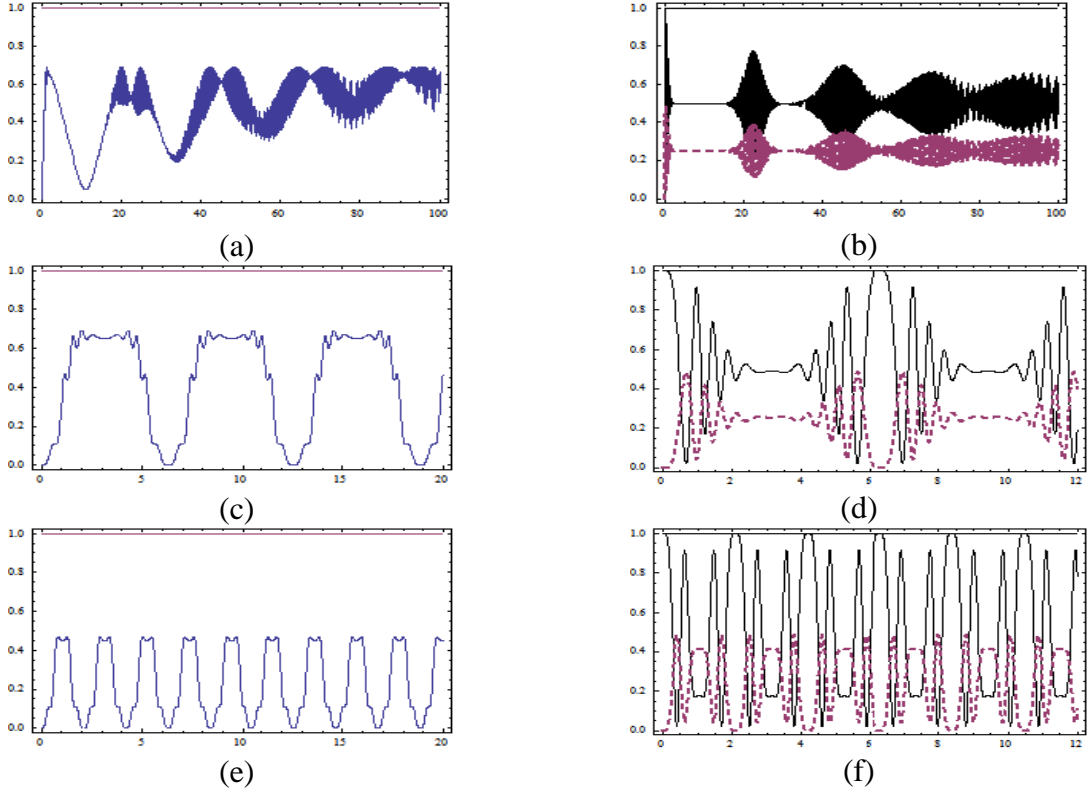

Fig.1.The time evolution of von Neumann entropy figures (a,c,e) and the population $\rho_{11}$ (solid line) and $\rho_{22}$ ( dashing line) figures (b,d,f) of the three- level system for $\bar{n} = 25$ and $\Delta_1 = \Delta_2 = 0$. Also, figures (a,b) the time atomic motion is neglected and the other figures atomic motion is considered $p = 1$ in figures (c,d) and $p = 3$ in figures (e,f)

Figure 1 corresponds to the entanglement due to von Neumann entropy and the populations in the absence of the detuning $\Delta_1$ and $\Delta_2$ (resonance case). It is remarkable that the entropy reaches minimum value. Also, the evolution of the entropy is periodic and the presence of the atomic motion leads to increase this periodically. In fig. 1 (b, d and f ), the time evaluation of the populations $\rho_{11}$ (solid line ) and $\rho_{11}$ (dash line) are given for the different value of $p$ in the resonance case. We notice that the collapse- revival phenomenon is very clear and the populations behave oscillator in the onset when $p$ is very small as shown in Fig.1 (b). On the other hand, the presence of atomic motion and the collapse and increase the oscillations and also increase the amplitude of oscillations as seen in Fig. 1(d). From Fig. 1 (f).

Figure 2 shows the influence of the detuning on the evolution of the entropy and on the population we put in the off-resonance ( $\Delta_1 = \Delta_2 = \Delta$ ) while the atomic motion takes the same values as in the previous figure. We observe from the figure, the amplitude of the field entropy is smaller than as in resonance case, and the oscillations are more periodic. Also, we notice that the field entropy reaches to its pure state for a long tome in the presence of the atomic motion. On the other hand, when we plot the populations with the same data, we see that the collapse and revival occur but the collapse time increases and the amplitude of the fluctuations decreases. Also, the atomic motion increases the periodically of the oscillations that shown in Fig. 2( b, d and f).

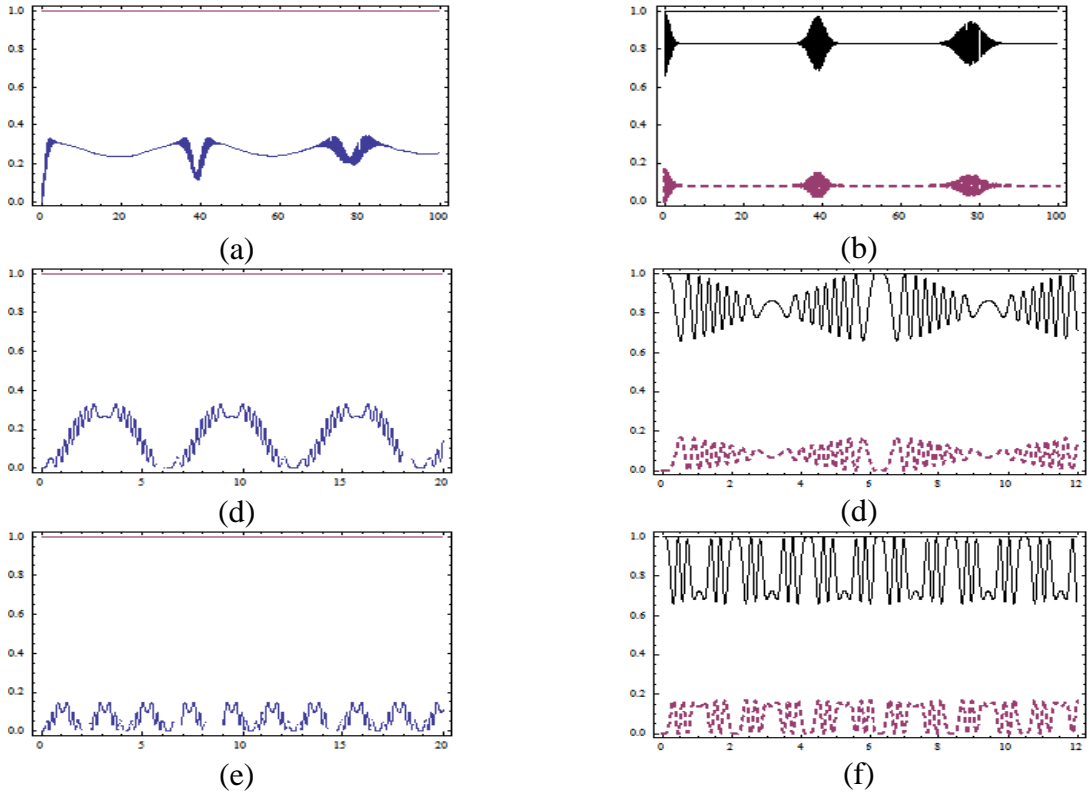

Fig. 2 the same as Fig. 1 but for $\Delta_1 = \Delta_2 = \Delta$

In figure 3, we consider the non-resonance case where we take $\Delta_1 = 0, \Delta_2 = 100$ the case of neglected atomic motion is plotted in Fig. 3(a) while the presences of atomic motion cases are plotted in Fig. 3 (c and e). The large detuning parameter the behavior of the considered system is similar of the two-level atom system.

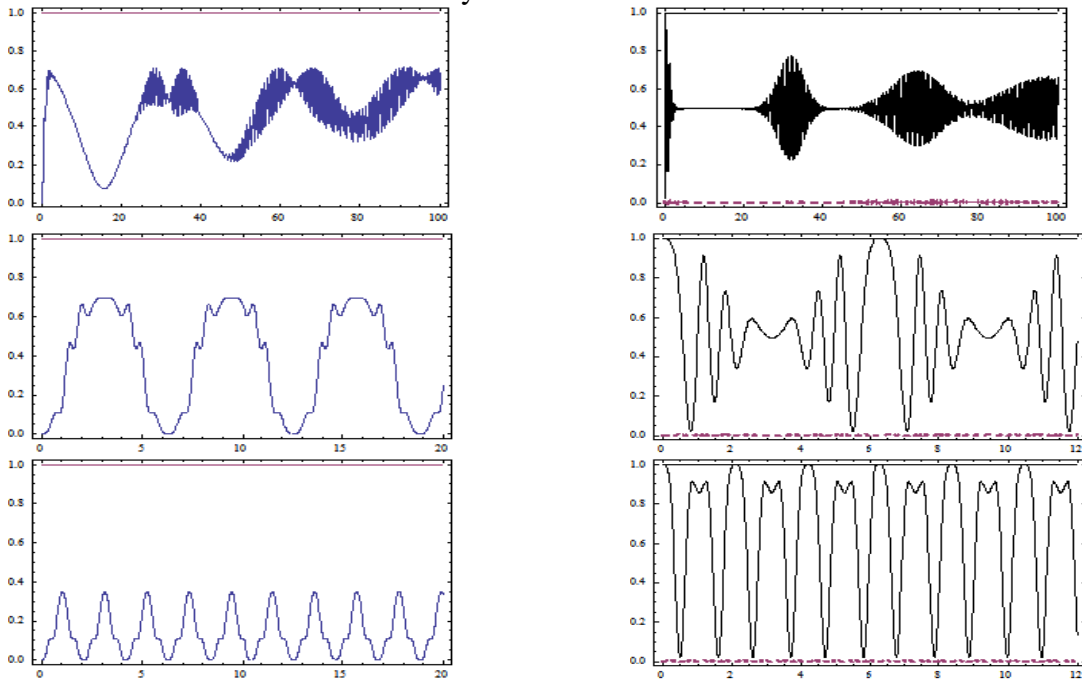

Fig. 3 the same as Fig. 1 but for $\Delta_1 = 0, \Delta_2 = 100$

We observe that the maximum entropy of the field subsystem that is achieved during the time evolution is inversely proportional to the detuning, that is, the bigger the detuning, the smaller the maximum entropy. Also, the fist maximum as the field entropy at t>0 is achieved at the collapse time. Furthermore, the entropy is periodic when the atom is in motion case. The time of a pure state for this case is small compared with the considered system; the amplitude of the oscillations also decreases as the parameter $p$ increases. On the other hand, the population show that the collapse- revival phenomenon for a long time when $\Delta_2$ is large. Also, we notice that the second collapse appears for weaken atomic motion. Moreover, considering of the atomic motion leads to increases the amplitude of the oscillation

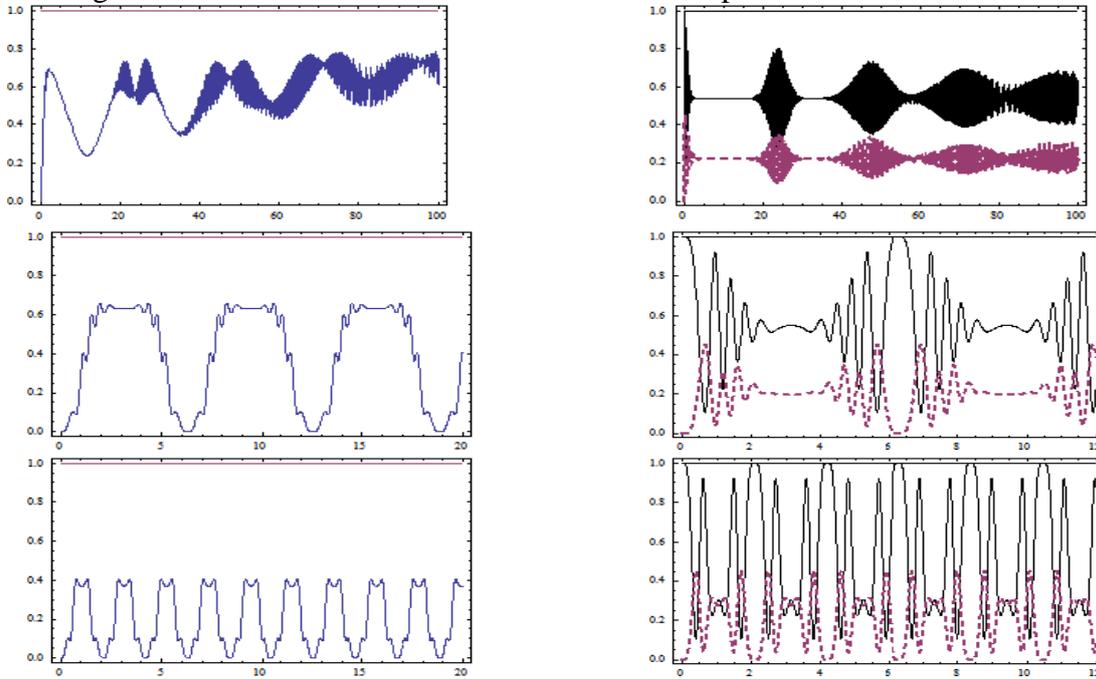

Fig. 4 the same as Fig. 1 but for $\Delta_1 = 5$ and $\Delta_2 = 4$

Figure 4 we consider the nonresonant case $\Delta_1 \neq \Delta_2$, and by a suitable choice of $\Omega$, $E_i$ we plot the von Neumann entropy and the populations $\rho_{11}$ (solid line) and $\rho_{11}$ (dash line) for the detunings $\Delta_1 = 5$ and $\Delta_2 = 4$ and the same different values of p as the above cases. We notice from this figure that the occurrence of collapse and revival depends upon the detunings and the entropy reaches to minimum value compared with the pervious case. Also, the evolution of entropy is periodic and the presence of the atomic motion leads to incensing this periodically. With respect to the population show clearly the collapse- revival phenomenon. The amplitude of oscillation and its periodically increase when the atomic motion is take in to account

## Conclusion

We have studied a three-level atom interacting with one-mode cavity field in the coherent state in the presence of the detuning parameters. Various statistical aspects are calculated and the population and the field entropy are investigated numerically for the atomic system of $\Lambda$-configuration. The effect of detuning parameters or/and the atomic motion on theses phenomena when the atom is initially prepared in the upper state are explored. We observe that the detuning parameters and the atomic motion have an important effect on the properties of the von Neumann entropy and the population.

## References

1. E. T. Jaynes and F. W. Cummings, *Proc. IEEE*, **51**, 89 (1963).
2. D. F. Walls and P. Zoller, Phys. Rev. Lett. **47**, 709 (1981).
3. K. Wodkiewicz, P.L. Knight, S. J. Buckle, and S. M. Barnett, Phys. Rev. A **35** 2567 (1987).

4. G. Hu and P. K. Aravind, J. Opt. Soc. Am. B **6**, 1757 (1989).
5. P. Zhou and J. S. Peng, Phys. Rev. A **44**, 3331 (1991).
6. M. Kozierowski, J. E. Poyatos, and L. L. Sanchez-Soto, Phys.Rev. A **51**, 2450 (1995).
7. Rui-hua Xie, Phys. Rev. A **53**, 2897 (1996).
8. C. C. Gerry and S. Rodrigues, Phys. Rev. A **36**, 5444 (1987).
9. C. C. Gerry and E. R. Vrscay, Phys. Rev. A **37**, 1779 (1988).
10. V. Buzek, Phys. Rev. A **39**, 5432 (1989).
11. P. Meystre and M. Sargent III, *Elements of Quantum Optics*-Springer-Verlag, New York
12. S. Feneuile, Rep. Prog. Phys. **40**, 1257 (1977).
13. X.-s. Li, D.L. Lin, C.D. Dong, Phys. Rev. A **36**, 5209 (1987).
14. Z.-D. Liu, X.-S. Li, D.L. Lin, Phys. Rev. A **36**, 5220 (1987).
15. N.H. Abdel-Wahab, Phys. Scr. **76**, 233 (2007).
16. N.H. Abdel-Wahab, Phys. Scr. **76**, 244 (2007).
17. P. Storey, M. Collet, D. Walls, Phys. Rev. Lett. **68**, 472 (1992).
18. P. Storey, M. Collet, D. Walls, Phys. Rev. A **47**, 405 (1993).
19. A. Vaglica, Phys. Rev. A **52**, 2319 (1994).
20. J. Liu, W. Wang, Phys. Rev. A **54**, 2326 (1996).
21. B. Yurke, D. Stoler, Phys. Rev. Lett. **57**, 13 (1993).
22. K. Zaheer, M.S. Zubairy, Phys. Rev. A **39**, 2000 (1989).
23. A.-S.F. Obada, A.M. Abdel-Hafez, Phys. Rev. A **43**, 5161 (1991).
24. M. F. Fang, *Physica A*, **204**, 193 (1994).
25. M. Sebawe Abdalla, A.-S. F. Obada, and S. Abdel-Khalek, *Chaos, Solitons, Fractals*, **36**, 405 (2008).
26. M. Abdel-Aty, *J. Phys. A: Math. Gen.*, **38**, 8589 (2005).
27. S. Abdel-Khalek, *Physica A*, **387**, 779 (2008).
28. Fang, M.F., Liu, H.E. Phys. Lett. A **200**, 250 (1995)
29. S.J.D. Phoenix and P.L. Knight, Ann. Phys. (NY) 186 381((1988).
30. S.J.D. Phoenix and P.L. Knight, Phys. Rev. A 44 6023(1991).
31. S.J.D. Phoenix and P.L. Knight, J. Opt. Sot. Am. B 7 116(1990).